\documentclass[twocolumn,preprintnumbers,showpacs,showkeys,amsmath,amssymb]{revtex4}
\usepackage{graphicx}
\usepackage{dcolumn}
\usepackage{bm}
\begin{document}
\title{Temporal evolution of product shock measures in TASEP with sublattice-parallel update}
\author{F. H. Jafarpour $^{1}$}
\email{farhad@ipm.ir}
\author{S. R. Masharian$^2$}
\affiliation{$^1$ Physics Department, Bu-Ali Sina University, 65174-4161 Hamedan, Iran}
\affiliation{$^2$ Institute for Advanced Studies in Basic Sciences, 45195-1159 Zanjan, Iran}
\date{\today}
\begin{abstract}
It is known that when the steady-state of a one-dimensional multi-species system,
which evolves via a random-sequential updating mechanism, is written in terms of a
linear combination of Bernoulli shock measures with random walk dynamics, it can be
equivalently expressed as a matrix-product state. In this case the quadratic algebra
of the system always has a two-dimensional matrix representation. Our investigations
show that this equivalence exists at least for the systems with deterministic
sublattice-parallel update. In this paper we consider the totally asymmetric simple
exclusion process on a finite lattice with open boundaries and sublattice-parallel update as an example.
\end{abstract}
\pacs{02.50.Ey, 05.20.-y, 05.70.Fh, 05.70.Ln}
\keywords{TASEP, Random walk, Shock measure, matrix-product approach}
\maketitle
\section{Introduction}
{\par Recently, the steady-state properties of exclusion processes which belong to the class of driven-diffusive
systems have been under much investigation because of their unique physical characteristics such as shocks and
non-equilibrium phase transitions \cite{Schm95}-\cite{Schu00}. The matrix-product approach has turned out to be
one of the most powerful techniques in determining the steady-states of such stochastic non-equilibrium systems
which have been used to model biological transport and traffic flow. The matrix-product approach has been interpreted
from different point of views (a recent review of this approach can be found in \cite{BE07}). According to this
approach the non-equilibrium steady-state weight of any given configuration of a one-dimensional stochastic system
can be considered as a matrix element of product of non-commuting operators, one for each lattice site, chosen
according to the state of the lattice site. In order to calculate these weights one needs to know a set of algebraic
relations between these operators. Whether these operators have matrix representations is a challenging issue.}
{\par We have recently investigated the relation between the dimensionality of the matrix representation of the
algebra of a given one-dimensional driven-diffusive system with nearest-neighbor interactions and the possibility
that the steady-state of the system in question can be written in terms of a linear superposition of product shock
measures. In this context a shock is defined as a sharp discontinuity in the density profile of particles on the
lattice. In \cite{JM07} we have shown that for the one-dimensional driven-diffusive systems defined on a discrete
lattice of finite size in which a single product shock measure has a simple random walk dynamics under the time evolution
generated by the stochastic Hamiltonian of the system, the steady-state of the system can easily be written as a
linear combination of these single product shock measures. In most cases it is  necessary that some constraints on
the microscopic reaction rates of the system are satisfied. Surprisingly we have seen that at the same time the
steady-state of the system can be written as a matrix-product state with two-dimensional matrix representation
for the algebra of the system, provided that the same constraints  (if they exist) on the microscopic reaction
rates are satisfied. The matrix representation in this case obeys an special structure which contains very
detailed information about the hopping rates of the shock front, as well as the density of particles on the
left and the right hand sides of the shock front. In present work we aim to investigate the same issue;
however, this time for the systems in a different updating scheme i.e. the sublattice-parallel dynamics.}
{\par One of the simplest, yet interesting, driven-diffusive model which has been studied widely during
recent years is the Asymmetric Simple Exclusion Process (ASEP). In this model the classical particles enter
the system from the left boundary of a discrete lattice, diffuse in its bulk and leave it from the right
boundary with certain rates. The derivation of the matrix-product representation from the algebraic Bethe
ansatz for this model has been studied in \cite{GM06}. For this model the equivalence between the partition
function of the system with random sequential dynamics and the partition function of a two-dimensional lattice
path model of one-transit walks or Dyck paths has been studied in \cite{BJJK04}. Under the parallel dynamics
the partition function of the ASEP can be expressed as one of several equivalent two-dimensional lattice path
models involving weighted Dyck paths\cite{BJJK07}. The dynamics of a single shock front in the ASEP with a
discrete time updating scheme defined on an infinite lattice has already been studied in \cite{BS00}.}
{\par In this paper we will answer the question whether the existence of a two-dimensional representation
for the quadratic algebra of a driven-diffusive system with a discrete time updating (more specifically
sublattice-parallel updating) and nearest-neighbor interactions is related to the fact that the steady-state
of the system can be constructed in terms of a linear superposition of product shock measures with simple
random walk dynamics. We will consider the Totally Asymmetric Simple Exclusion Process known as TASEP with
open boundaries as a simple example. In this model the particles only hop towards the right boundary after
being injected into the lattice from the left boundary. As we will see, quite similar to the case of random
sequential updating scheme studied in \cite{JM07}, it seems that whenever the quadratic algebra of the system
(since we only consider the systems with nearest-neighbors interactions) has a two-dimensional matrix
representation with a specific structure (which will be discussed later) then we can conclude that the
steady-state of the system is made up of a linear combination of product shock measures with a shock front
which has simple random walk dynamics and vice versa. One of the differences here is that in the case of
sublattice updating scheme one should define two different shocks which behave differently at even or odd
lattice sites. This has already been observed in \cite{BS00} for an infinite system.}
{\par In the following sections we will first present the mathematical tools and definitions. Then we will
study the time evolution of two product shock measures defined at even and odd sites under the
sublattice-parallel update. Then we will construct the steady-state of the system in terms of a
linear combination of these shocks. We will bring the quadratic algebra of the system and its
two-dimensional representation in terms of the shock characteristics. The conclusion will be
presented in the last section.}
\section{Mathematical preliminaries}
{\par Let us start with the definitions and mathematical preliminaries. Consider a general
two-state driven-diffusive system with nearest-neighbor interactions and sublattice-parallel
dynamics in which classical particles move on a one-dimensional lattice of length $2L$ with
open boundaries. The bulk dynamics consists of two half time steps. In the first half time
step even lattice sites and also the first and the last lattice sites are updated. From the
first and the last lattice sites the particles can be injected or extracted with certain
probabilities. In the second half time step only the odd lattice sites are updated. The
corresponding transfer matrix $T$ consists of two factors $T=T_2 T_1$ defined as \cite{HH96}:
\begin{eqnarray}
\label{T1}
T_1 &=& {\cal L} \otimes {\cal T} \otimes \ldots \otimes
        {\cal T} \otimes {\cal R} \,\;=\;\,
{\cal L} \otimes {\cal T}^{\otimes (L-1)} \otimes {\cal R} \nonumber\\
\label{T2}
T_2 &=& \hspace{3.8mm} {\cal T} \otimes {\cal T} \otimes \ldots \otimes
        {\cal T} \hspace{3.8mm} \,\;=\;\,  {\cal T}^{\otimes L} \nonumber
\end{eqnarray}
where ${\cal T}$, ${\cal L}$ and ${\cal R}$ are the matrices for bulk interactions, particle input and particle output respectively.
The time evolution of the probability distribution is governed by the following equation:
\begin{equation}
\label{Time Evol}
T \vert P(t) \rangle
= \vert P(t+1) \rangle.
\end{equation}
In long-time limit the system approaches its steady-state and its non-equilibrium probability distribution
satisfies the following equation:
\begin{equation}
\label{SS}
T \vert P^{\ast} \rangle = \vert P^{\ast} \rangle.
\end{equation}}
{\par As a simple example consider the TASEP which in an appropriate basis the above transfer matrix can be written as \cite{HH96}:
\begin{widetext}
\begin{equation}
\label{Interaction}
{\cal T} \;=\; \left(
\begin{array}{cccc}
1 & 0 & 0 & 0 \\
0 & 1 & 1 & 0 \\
0 & 0 & 0 & 0 \\
0 & 0 & 0 & 1
\end{array} \right),\;
{\cal L} \;=\; \left(
\begin{array}{cc}
1-\alpha & 0 \\
\alpha & 1
\end{array} \right),\;
{\cal R} \;=\; \left(
\begin{array}{cc}
1 & \beta \\
0 & 1-\beta
\end{array} \right) .
\end{equation}
\end{widetext}
As can be seen, the particles in the bulk of the lattice move only to the right deterministically while
obeying the exclusion principle. The particles can enter (leave) the lattice only from the left (right)
boundary with the probability $\alpha$ ($\beta$). The dynamics and also the steady-state of the TASEP with
open boundaries has been proposed and studied in \cite{SCH93}. Later its steady-state was studied using a
matrix formalism in \cite{HH96}. In the thermodynamic limit i.e. $L>>1$ one finds that the system has two
different phases: a high-density phase for $\alpha > \beta$ and a low-density phase for $\alpha < \beta$.
A first-order phase transition also takes place at the transition line $\alpha=\beta$.}
\section{Temporal evolution of shocks}
{\par In what follows we study the time evolution of two product shock measures using the time evolution
equation (\ref{Time Evol}). We consider a discrete lattice of length $2L$ and introduce two product shock
measures at even sites $2k$ ($k=1,\cdots,L$) as $\vert \mu_{2k}\rangle$ and at odd sites $2k+1$ ($k=0,\cdots,L$)
as $\vert \mu_{2k+1} \rangle$ according to:
\begin{widetext}
\begin{eqnarray}
\vert \mu_{2k} \rangle=
  \underbrace{\left(\begin{array}{c}
  1-\rho_{1}^{odd} \\ \rho_{1}^{odd}
  \end{array}\right)}_{1}\otimes
  \left(\begin{array}{c}
  1-\rho_{1}^{even} \\ \rho_{1}^{even}
  \end{array}\right)\otimes \cdots \otimes
 \underbrace{ \left(\begin{array}{c}
  1-\rho_{1}^{odd} \\ \rho_{1}^{odd}
  \end{array}\right)}_{2k-1}\otimes
 \underbrace{ \left(\begin{array}{c}
  1-\rho_{2}^{even} \\ \rho_{2}^{even}
  \end{array}\right)}_{2k}\otimes \cdots \otimes
  \left(\begin{array}{c}
  1-\rho_{2}^{odd} \\ \rho_{2}^{odd}
  \end{array}\right)\otimes
  \underbrace{\left(\begin{array}{c}
  1-\rho_{2}^{even} \\ \rho_{2}^{even}
  \end{array}\right)}_{2L}\\
  \vert \mu_{2k+1} \rangle=
  \underbrace{\left(\begin{array}{c}
  1-\rho_{1}^{odd} \\ \rho_{1}^{odd}
  \end{array}\right)}_{1}\otimes
  \left(\begin{array}{c}
  1-\rho_{1}^{even} \\ \rho_{1}^{even}
  \end{array}\right)\otimes \cdots \otimes
 \underbrace{ \left(\begin{array}{c}
  1-\rho_{1}^{even} \\ \rho_{1}^{even}
  \end{array}\right)}_{2k}\otimes
 \underbrace{ \left(\begin{array}{c}
  1-\rho_{2}^{odd} \\ \rho_{2}^{odd}
  \end{array}\right)}_{2k+1}\otimes \cdots \otimes
  \left(\begin{array}{c}
  1-\rho_{2}^{odd} \\ \rho_{2}^{odd}
  \end{array}\right)\otimes
  \underbrace{\left(\begin{array}{c}
  1-\rho_{2}^{even} \\ \rho_{2}^{even}
  \end{array}\right)}_{2L}
\end{eqnarray}
\end{widetext}}
{\par We should explain a couple of points here. Firstly, one should note that the shock front for
$\vert \mu_{2k}\rangle$ lies between the lattice sites $2k-1$ and $2k$ while the shock front for
$\vert \mu_{2k+1}\rangle$ lies between the lattice sites $2k$ and $2k+1$. Secondly, the shock
$\vert \mu_{2L+1}\rangle$ indicates a flat distribution of particles with densities $\rho_{1}^{odd}$
and $\rho_{1}^{even}$ at odd and even lattice sites respectively. In this case the shock front can be
considered to be between the lattice sites $2L$ and $2L+1$ which the later is considered as an auxiliary
site. Let us consider $\vert \mu_{2k} \rangle$ and $\vert \mu_{2k+1} \rangle$ as two initial probability
distributions and investigate their time evolutions using (\ref{Time Evol}). Suppose that under some
constraints on the reaction rates the dynamics of these two shocks are given by:
\begin{widetext}
\begin{equation}
\label{Eq}
\begin{array}{l}
T\vert \mu_{2k} \rangle=\delta_{l} \vert \mu_{2k-1} \rangle+\delta_{r} \vert \mu_{2k+1} \rangle+\delta_{s} \vert \mu_{2k} \rangle \;\; \mbox{for}\;\; 1 \leq k \leq L\\ \\
T\vert \mu_{2k+1} \rangle=\delta_{l} \delta_{s} \vert \mu_{2k} \rangle +\delta_{r} \delta_{s} \vert \mu_{2k+2} \rangle + \delta_{r}^{2} \vert \mu_{2k+3} \rangle+\delta_{l}^2\vert \mu_{2k-1} \rangle+(\delta_{s}+2\delta_{l}\delta_{r})\vert \mu_{2k+1} \rangle\;\; \mbox{for} \;\; 1 \leq k \leq L-1\\ \\
T\vert \mu_{1} \rangle=\delta_{r} \delta_{s} \vert \mu_{2} \rangle+\delta_{r}^2 \vert \mu_{3} \rangle+(1-\delta_{r}\delta_{s}-\delta_{r}^2) \vert \mu_{1} \rangle \\ \\
T\vert \mu_{2L+1} \rangle=\delta_{l} \delta_{s} \vert \mu_{2L} \rangle+\delta_{l}^2 \vert \mu_{2L-1} \rangle+(1-\delta_{l}\delta_{s}-\delta_{l}^2) \vert \mu_{2L+1} \rangle
\end{array}
\end{equation}
\end{widetext}
in which $\delta_{s}=1-\delta_{l}-\delta_{r}$.}
{\par The first two equations in (\ref{Eq}) have already been obtained in \cite{BS00} for the ASEP on an infinite lattice;
however, since the definition of the transfer matrix in this paper is based on \cite{HH96}, these two time evolution
equations have exchanged places as compared to the corresponding equations (44) and (48) in \cite{BS00}. For the TASEP
with open boundaries these equations of motion are valid, provided that we have:
\begin{equation}
\label{TASEP}
\begin{array}{l}
\rho_1^{odd}=0 \; , \; \rho_2^{odd}=1-\beta\\
\rho_1^{even}=\alpha \; , \; \rho_2^{even}=1\\
\delta_{r}=\beta \; , \; \delta_{l}=\alpha.
\end{array}
\end{equation}
Note that (\ref{Eq}) gives a closed set of time evolution equations for $\vert \mu_k\rangle$'s in which $k=1,\cdots,2L+1$.
It is also interesting to note that in contrast to the ASEP, here there is no constraint on the microscopic reaction rates
(the boundary rates $\alpha$ and $\beta$). Although the particles move deterministically towards the right boundary, the shock
fronts hop both to the left and to the right. This is a direct result of the updating scheme.}
\section{Steady-state}
{\par The simplicity of the time evolution equations (\ref{Eq}) allows us to construct the steady-state of the system
$\vert P^{\ast} \rangle$. As can be seen, they are similar to the time evolution equations for a simple random walker
moving on a finite lattice with reflecting boundaries; however, the random walker behaves differently when it lies at
an even or an odd lattice site. By considering a linear superposition of the shocks as:
\begin{equation}
\label{Distro}
\vert P^{\ast} \rangle = \frac{1}{Z}\sum_{k=1}^{2L+1}c_k \vert \mu_k \rangle
\end{equation}
and requiring that (\ref{SS}) should be satisfied one finds:
\begin{eqnarray}
&& c_{2k}=\delta_{s}(\frac{\delta_{r}}{\delta_{l}})^{2k-1} \; \mbox{for} \; k=1,\cdots,L\\
&& c_{2k+1}=(\frac{\delta_{r}}{\delta_{l}})^{2k} \; \mbox{for} \; k=0,\cdots,L.
\end{eqnarray}
The normalization factor or the partition function of the system $Z$ can be easily calculated as:
\begin{equation}
\begin{array}{ll}
Z&=\sum_{k=1}^{2L+1}c_k\\ & \\
&=\frac{1}{\delta_{l}-\delta_{r}}(\delta_{l}(1-\delta_{r})-\delta_{r}(1-\delta_{l})(\frac{\delta_{r}}{\delta_{l}})^{2L}).
\end{array}
\end{equation}
Since the steady-state of the system is unique, if one calculates the steady-state probability distribution of the
system using the matrix-product approach, one should find the same distribution as (\ref{Distro}).}
\section{Matrix product approach}
{\par Let us now investigate the steady-state probability distribution function of our general two-state model
with sublattice-parallel dynamics and nearest-neighbor interactions using the matrix-product approach. According
to this approach (and in this particular updating scheme) the steady-state probability distribution function of
the system can be written as \cite{HH96}:
\begin{equation}
\label{MatrixAnsatz}
\vert P^{\ast}\rangle= \frac{1}{Z} \langle \langle W \vert
\left[\left( \begin{array}{c}
{\hat E} \\ {\hat D} \end{array} \right) \otimes \left( \begin{array}{c}
E \\ D \end{array} \right)\right]^{\otimes L}
\vert V \rangle \rangle
\end{equation}
in which the operators $\hat E$ and $\hat D$ ($E$ and $D$ ) stand for the presence of a hole and a particle at
odd (even) sites respectively. The normalization factor $Z$ is usually called the partition function and can
easily be written in a grand canonical ensemble as:
\begin{equation}
Z=\langle \langle W \vert ({\hat E}+{\hat D})^L(E+D)^L \vert V\rangle \rangle.
\end{equation}
The operators ($\hat{E},\hat{D}$) and ($E,D$) besides the vectors $\vert V \rangle\rangle$ and
$\langle \langle W \vert$ are acting in an auxiliary space. According to the standard matrix-product
approach by requiring that (\ref{SS}) is satisfied one finds that the above mentioned operators and
vectors should satisfy a quadratic algebra given by \cite{HH96}:
\def\edmatrix{\Bigl(
    \begin{array}{c} E \\ D
    \end{array}\Bigr)}
\def\abmatrix{\Bigl(
    \begin{array}{c} \hat{E} \\ \hat{D}
    \end{array} \Bigr)}
\begin{eqnarray}
&&{\cal T} \, \left[ \edmatrix \otimes \abmatrix \right] \;=\;\abmatrix \otimes \edmatrix  \nonumber\\
&&\langle\langle W \vert {\cal L} \abmatrix \;=\; \langle \langle W \vert \edmatrix\\
&&{\cal R} \edmatrix \vert V \rangle\rangle\;=\; \abmatrix\vert V \rangle\rangle  \nonumber.
\end{eqnarray}
Surprisingly, one can see that the following two-dimensional matrix representation which can be written
in terms of the shock hopping rates and the densities of the Bernoulli measures at the left and the right
hand sides of the shock position can generate exactly the same probability distribution (\ref{Distro}):
\begin{widetext}
\begin{equation}
\begin{array}{l}
\label{BulkRep}
\hat{D} = \left( \begin{array}{cccc}
\rho_{2}^{odd}& & & 0 \\
\hat{d} & & & \frac{\delta_{r}}{\delta_{l}}\rho_{1}^{odd}
\end{array} \right) \; , \;
\hat{E} = \left( \begin{array}{cccc}
1-\rho_{2}^{odd}& & & 0 \\
-\hat{d} & & & \frac{\delta_{r}}{\delta_{l}}(1-\rho_{1}^{odd})
\end{array} \right) \; ,\; \\ \\
D = \left( \begin{array}{cccc}
\rho_{2}^{even}  & & & 0 \\
d & & & \frac{\delta_{r}}{\delta_{l}}\rho_{1}^{even}
\end{array} \right) \; , \;
E = \left( \begin{array}{cccc}
1-\rho_{2}^{even} & & & 0  \\
-d & & & \frac{\delta_{r}}{\delta_{l}}(1-\rho_{1}^{even})
\end{array} \right) \; , \; \\ \\
\langle\langle W \vert = (w_1, w_2) \; , \;
\vert V \rangle \rangle= \left( \begin{array}{c}
v_1 \\ v_2 \end{array} \right)
\end{array}
\end{equation}
\end{widetext}
provided that we have:
\begin{equation}
\left\{
\begin{array}{ll}
v_1w_2d=\frac{(\delta_{r}-1)(\rho_{2}^{even}-\rho_{1}^{even})}{\frac{\delta_{l}}{\delta_{r}}-1},\\
v_1w_2\hat{d}=\frac{(\delta_{l}-1)(\rho_{2}^{odd}-\rho_{1}^{odd})}{\frac{\delta_{l}}{\delta_{r}}-1}
\end{array}.
\right.
\end{equation}
These relations are nothing but two constraints on the parameters $v_1$, $v_2$, $w_1$, $w_2$, $d$ and $\hat d$;
therefore, only four of these parameters are free. Note that the densities in the shock measures and also the
shock front hopping rates in (\ref{Eq}) should be fixed by the boundaries and the microscopic reaction rates;
therefore, are not free parameters.}
{\par Let us go back to our simple example. It is shown in \cite{HH96} that the TASEP has a quadratic algebra
which can be written as:
\begin{widetext}
\begin{equation}
\begin{array}{l}
\label{BulkAlgebra}
[E,\hat{E}] = [D,\hat{D}] = 0 \; ,
\; E \hat{D} = [\hat{E},D] \; , \; \hat{D} E = 0 \; , \;
\langle\langle W \vert \hat{E} (1-\alpha) = \langle \langle W \vert E \\
\langle\langle W \vert (\alpha \hat{E} + \hat{D}) = \langle \langle W \vert D \; , \;
(1-\beta) D \vert V \rangle\rangle = \hat{D} \vert V \rangle\rangle \; , \;
(E+\beta D) \vert V \rangle \rangle= \hat{E}\vert V \rangle\rangle.
\end{array}
\end{equation}
\end{widetext}
In the same reference it has been shown that (\ref{BulkAlgebra}) has a two-dimensional representation
for $\alpha \neq \beta$ which can be simply shown that it is of the form (\ref{BulkRep}) with the
parameters given in (\ref{TASEP}).}
{\par The reason that we emphasis the matrix representation of the algebra (\ref{BulkAlgebra}) can be
rewritten in the form of (\ref{BulkRep}) (which is slightly different from what was first proposed in
\cite{HH96}) is as follows: As we have claimed in our previous papers, whenever the steady-state of a
one-dimensional driven-diffusive system defined on a finite or infinite open lattice which evolves
under the random sequential updating scheme can be written in terms of a linear superposition of
Bernoulli shocks with simple random walk dynamics, then the algebraic relation between the operators
(when the steady-state is studied using the matrix-product formalism) will have a two-dimensional
representation with a generic structure. In \cite{JM07} we have also proposed a general formalism
by which one can simply find a two-dimensional representation for the quadratic algebra of the system
in terms of the hopping rates of the shock front and the densities of the particles on the left and the
right hand sides of the shock. This works if and only if the time evolution of the position of a product
shock measure with a single shock front is simply a random walk. Moreover it has been shown, by providing
several examples, that the conditions under which the domain wall has a random walk dynamics are exactly
those for the existence of the two-dimensional matrix representation \cite{JM07}.}
\section{Concluding remarks}
{\par Let us review the results of the current work. The most important goal in this work was answering to
the question whether the matrix representation of the quadratic algebra of the system developing vis
sublattice-parallel updating scheme has the same generic structure as we had proposed for the case of
continuous-time updating scheme? By comparing our results in this paper with those in \cite{JM07} one
finds that the matrix representation retains its structure even in sublattice-parallel updating scheme
and it seems that, although we have no direct proof for it at the moment, the same is true for other
updating schemes. In this direction we have considered a general driven-diffusive system with nearest-neighbor
interactions which evolve vis sublattice-parallel updating scheme and is defined on an open lattice. If we assume
that a single product shock measure has a simple random walk dynamics, generated by the transfer matrix of the system,
then the steady-state of this system can easily be written in terms of a linear superpositions these shock measures.
On the other hand we have introduced a two-dimensional matrix representation which can generate exactly the same steady-state.
The nontrivial point is that this matrix representation has exactly the same structure that we had found for the case of
continuous time updating scheme.\\
As an evidence we have studied the TASEP under the sublattice-parallel updating scheme and shown that an uncorrelated
shock can evolve in the system without requiring any constraints on the microscopic reaction rates i.e. the injection
and the extraction rates of the particles. The shock also reflects from the boundaries of the lattice with some nonzero
rates. By investigating the time evolution equations of the shock front we have found that it has simple random walk dynamics.
Since the dynamics of the shock front is quite similar to that of a random walker, the steady-state of the system can be
constructed as a linear superposition of such product shock distributions. This could have been supposed since our
experience with random sequential updating scheme had shown that in this case the quadratic algebra of the system
should have a two-dimensional matrix representation as it was found in \cite{HH96}. As we have seen in this paper a
two-dimensional matrix representation for the quadratic algebra of the TASEP under discrete time updating exists
without any constraints.\\
Our investigations show that the ASEP with the most general 4-parameter open boundary conditions studied in
\cite{HP97}, can be explained using our approach provided that the same conditions under which the quadratic
algebra of the system has a two-dimensional matrix representation are fulfilled. We have also found completely
new families of driven-diffusive models evolving under sublattice-parallel updating scheme in which a product
shock measure with a single shock front has a simple random walk dynamics very similar to the equations (\ref{Eq}),
provided that some constraints on the microscopic reaction probabilities are satisfied. We have shown that the
steady-state of these systems can be written in terms of a combinations of such single shocks and at the same
time the matrix representation of the quadratic algebras of these systems has the same unique structure as
in (\ref{BulkRep}). The details of these results will be presented elsewhere.}
\section*{Acknowledgment}
The authors would like to thank V. Popkov for reading the manuscript and giving enlightening comments.

\end{document}